# Rheological and Physicochemical Studies on Emulsions Formulated with Chitosan Previously Dispersed in Aqueous Solutions of Lactic Acid

Lucas de Souza Soares[1], Janaína Teles de Faria[1,2], Matheus Lopes Amorim[1], João Marcos de Araújo[3], Luis Antonio Minim[1], Jane Sélia dos Reis Coimbra[1], Alvaro Vianna Novaes de Carvalho Teixeira[4], Eduardo Basílio de Oliveira[1].

**Abstract** Chitosan, a natural, cationic polysaccharide, may be a hydrocolloid strategic to formulate acidic food products, as it can act as both bio-functional and technofunctional constituent. Typically, acetic acid is used to disperse chitosan in aqueous media, but the use of this acid is limited in food formulations due to its flavor. In this study, chitosan was firstly dispersed (0.1% m/V) in lactic acid aqueous solutions (pH 3.0, 3.5 or 4.0), and then evaluated regarding its thickener and emulsion stabilizer properties. O/W emulsions were prepared and characterized in terms of rheological properties, droplets average diameters and droplets ζ-potential. Emulsions containing chitosan were 3 times more viscous than controls without chitosan, and presented storage modulus ($G'$) higher than loss modulus ($G''$). Furthermore, they displayed two different populations of droplets (average diameters of 44 and 365 nm) and positive ζ-potential values (+50 mV). Droplets average diameters and ζ-potential did not present significant changes ($p > 0.05$) after storage at 25 °C during 7 days. This study showed that i) food organic acids other than acetic acid can be used to disperse chitosan for technological purposes, and ii) chitosan dispersed at very low concentrations (0.1% m/V) had relevant effects on rheological and physicochemical aspects of food-grade emulsions.

**Abbreviations**

| | |
|---|---|
| $a$ | Second constant of Mark-Houwink-Sakurada relationship (dimensionless) |
| $A$ | Fitting constant in equation (4) (mN·m$^{-1}$) |
| $b$ | Constant related to the interfacial tension decay rate in equation (4) (s$^{-0.5}$) |
| $C$ | First constant of Mark-Houwink-Sakurada relationship (dL·g$^{-1}$) |
| CI | Creaming index (%) |
| $d$ | Average diameter of emulsion droplets (nm) |
| DD | Deacetylation degree (%) |
| DLS | Dynamic light scattering |
| $G'$ | Storage modulus (Pa) |
| $G''$ | Loss modulus (Pa) |
| $K$ | Consistency index (Pa·s$^n$) |
| MAPE | Mean absolute percentage error (%) |
| $M_v$ | Viscometric molar mass (kDa) |
| $n$ | Flow behavior index (dimensionless) |
| $R^2$ | Coefficient of determination (dimensionless) |
| $t$ | Time in equation (4) (s) |
| $V_C$ | Creamed oil volume (mL) |
| $V_T$ | Total emulsion volume (mL) |
| $\zeta$ | Zeta potential (mV) |
| $\mu$ | Viscosity (Pa·s) |
| $\sigma$ | Interfacial tension (mN·m$^{-1}$) |
| $\sigma_{eq}$ | Interfacial tension at the equilibrium in equation (4) |

✉ Eduardo Basílio de Oliveira eduardo.basilio@ufv.br

[1] Departamento de Tecnologia de Alimentos (DTA), Universidade Federal de Viçosa (UFV), Viçosa, MG CEP 36570-900, Brazil.

[2] Instituto de Ciências Agrárias (ICA), Universidade Federal de Minas Gerais (UFMG), Montes Claros, MG CEP 39400-000, Brazil.

[3] Departamento de Biologia Geral (DBG), Universidade Federal de Viçosa (UFV), Viçosa, MG CEP 36570-900, Brazil.

[4] Departamento de Física (DPF), Universidade Federal de Viçosa (UFV), Viçosa, MG CEP 36570-900, Brazil.

|     | (mN·m$^{-1}$) |
| --- | --- |
| $\tau$ | Shear stress (Pa) |
| $\omega$ | Frequency (Hz) |
| $\dot{\gamma}$ | Shear rate (s$^{-1}$) |
| $[\eta]_H$ | Huggins intrinsic viscosity (dL·g$^{-1}$) |
| $[\eta]_K$ | Kraemer intrinsic viscosity (dL·g$^{-1}$) |
| $\overline{[\eta]}$ | Average between Huggins and Kraemer intrinsic viscosities (dL·g$^{-1}$) |

## Introduction

Chitosan [poly-β(1→4)-D-glucosamine] is a polysaccharide obtained from chitin extracted from shrimps and crab shells processed by food industries [1]. This biopolymer has a wide and well-known range of biotechnological applications, including material for bone and dermal prostheses, excipient for controlling release of bioactive compounds, bio-based films, and supports for enzyme immobilization [2–6]. Also, great interest is currently devoted to chitosan by both scientific community and the nutraceutical industry because of the hypolipidemic properties of this biopolymer. In the small intestine, chitosan interacts with bile salts and acids, forming aggregates which involve substances like cholesterol, triglycerides and free fatty acids [7,8][9]. Consequently, an important fraction of these substances can be excreted without being metabolized. Based on these physiological effects, there is nowadays a considerable variety of commercial products consisting simply of powdered chitosan in capsules (sometimes combined with vegetal fibers), with claims such as "blood cholesterol levels lowering" or even "slimming". Then, the use of chitosan to formulate acidic fluid foods, such as emulsion-based sauces and similar products, is an attractive idea.

Because of its above mentioned abilities to restrict the bioavailability of lipids in the gastrointestinal tract, chitosan could be envisioned as a bio-functional component and, simultaneously, as a thickener/stabilizer agent in emulsified formulations, replacing other polysaccharides commonly used for this purpose (carboxymethylcellulose, locust bean gum, modified starches, etc.) [10,11]. Indeed, some literature reports have addressed the physicochemical and techno-functional effects of adding chitosan to emulsions. For instance, Calero, Muñoz, Cox, Heuer, & Guerrero [12] studied O/W emulsions containing potato protein as emulsifier, and chitosan at up to 1.0 g·(100 g)$^{-1}$ previously dispersed in acetic acid/sodium acetate buffer. Kaasgaard & Keller [13] investigated similar O/W emulsions, with chitosan concentrations also reaching 1.0 g·(100 g)$^{-1}$, but using the negatively charged emulsifier CITREM LR10, in order to examine whether combinations of emulsifier and polysaccharide contribute to form emulsions and to control the instability of these systems. Klinkesorn & Namatsila [14] worked with concentrations of chitosan as high as 10.0 g·(100 g)$^{-1}$ (previously dispersed in 10 mM acetic acid solution at pH 6.0), and used the neutral emulsifier Tween 80 to form the emulsions. In all of these cases, acetic acid and/or acetates were used to disperse the chitosan prior to the preparation of emulsions, and high concentrations of chitosan were employed. However, the use of acetic acid may be unsuitable for some food applications, due to sensory limitations imposed by this acid. Hence, a proper characterization of emulsion systems containing chitosan, but dispersed using other food grade acids such as lactic acid, becomes necessary in a context of food product development.

In this paper, O/W emulsions were prepared using sunflower oil and aqueous solutions of lactic acidic containing dispersed chitosan, at a concentration lower than those found in literature [0.1 g·(100 g)$^{-1}$]. The electrically neutral food-grade emulsifier Tween 20 was chosen to be used in the study. Rheological and physicochemical analyses were undertaken to evaluate the effects of adding chitosan on different characteristics of the emulsions, compared with analogous systems but without chitosan.

## Experimental

### Materials

Chitosan was bought from Sigma-Aldrich Corporation (The United States of America; product ID =448,877). Before us- age, chitosan was abundantly washed with deionized water (QUV3, Millipore, Italy; 18.2 MΩ.cm, 25 °C) and lyophilized (LS 3000, Terroni, Brazil). Glacial acetic acid and sodium acetate (Vetec, Brazil; purity = 99.7%) were used to prepare chitosan dispersions for capilar viscosimetry characterization. Sodium azide (Labsynth, Brazil; purity = 99%), lactic acid (Impex Quimica, Spain; purity = 85%), Tween 20 (Sigma- Aldrich Corporation, The United States of America; product ID = P1379) and deionized water were used to prepare the acidic chitosan dispersions to form continuous phases during emulsification processes. Disperse phases were constituted of sunflower oil (Bunge, Brazil) colored with Sudan Black B (Dinâmica Química Contemporânea, Brazil; purity = 99%). All these chemicals are of analytical grade and were used as bought without further purification.

### Preliminary Chitosan Characterization

Viscometric molar mass ($M_v$) and deacetylation degree ($DD$) of the chitosan sample used in the present study were evaluated, according to the procedures followed by Amorim et al. [15]. Briefly, flow times of the diluted chitosan dispersions [0.1, 0.2, 0.3, 0.4, and 0.5 g·(100 mL)$^{-1}$] in acetic acid-sodium acetate buffer were measured (Cannon-Fenske viscometer 513 10, Schott, Germany). The Huggins and

Kraemer intrinsic viscosities were calculated ($[\eta]_H$ = 7.1 dL·g$^{-1}$ and $[\eta]_K$ = 8.4 dL·g$^{-1}$, respectively), giving an average intrinsic viscosity equal to $\overline{[\eta]}$ = 7.8 dL·g$^{-1}$. The constants of Mark-Houwink-Sakurada relationship ($\overline{[\eta]} = CM_v^a$) were determined according to Kasaai [16], and were $a$ = 0.93 and $C$ = 3.63·10$^{-5}$ dL·g$^{-1}$. Thus, the viscometric average molecular mass of chitosan was estimated as $M_v$ = 540 kDa. Deacetylation degree was estimated using a Raman spectroscopy approach [17], using a 110/S Bruker (Germany) apparatus. The integral intensities of the bands corresponding to wavenumbers 896 cm$^{-1}$, 936 cm$^{-1}$, 1591 cm$^{-1}$, and 1654 cm$^{-1}$ in the Raman spectrum of chitosan (Supplementary Material) were calculated through deconvolution in Lorentzian components [17, 18]. From these integral intensity values, the *DD* of the chitosan sample was estimated as (84.6 ± 5.1) %.

**Preparation of O/W Emulsions**

Firstly, lactic acid solutions were prepared by dropping 0.1mMlactic acid solution in predetermined amounts of water containing sodium azide [0.003 g·(100 mL)$^{-1}$], until reaching the desired pH value (3.0, 3.5, or 4.0), monitored using a digital pHmeter (H2221, Hanna, The United States of America). Thus, 0.1 g·(100 mL)$^{-1}$ of chitosan was added and the resulting systems were kept under stirring at 25.0 ± 0.1 °C in a thermostatic bath (TE-184, Tecnal, Brazil), during 24 h. After this, the pH of chitosan dispersions was again adjusted to their respective initial values. Finally, these dispersions were filtered using a 0.45 mm cellulose acetate membrane (Millipore, USA), in order to remove any non-dispersed material.

Acid lactic solutions or chitosan dispersions in acid lactic solutions (pH 3.0, 3.5, or 4.0) were added of the emulsifier Tween 20 [1.0 mL·(100 mL)$^{-1}$], and the resulting mixtures were used as continuous phases for preparing oil-in-water (O/W) emulsions. Emulsions were prepared with 90.0 mL of aqueous phase and 10.0 mL of colored sunflower oil. The emulsification process was started with the dispersion of oil within the aqueous phase using an agitator (Ultra Turrax DI 25 Basic, Yellow Line, India) at 24,000 rpm during 1 min, followed by the homogenization of the mixture with six passes through a high-pressure homogenizer (Emulsiflex-C5, Avestin, Canada), under pressure of 69 MPa. The obtained emulsions with different pH were respectively coded as A30, A35, and A40 (without chitosan), and Q30, Q35, and Q40 (with chitosan). They were transferred to graduated glass test tubes (12 mm internal diameter and 125 mm height), which were tightly sealed with a plastic film (to avoid air exchanges) and covered with aluminum foil (to protect against light pro-oxidative effects), and then stored at room temperature (25.0 ± 1.0 °C).

**Characterization of Emulsions**

Rheological analyses were performed immediately after preparation of emulsions ($t$ = 0). Average droplet size, $\zeta$-potential of droplets, and creaming analyses were carried out immediately after preparation ($t$ = 0) and after seven days of storage ($t$ = 7).

*Rheological Analyses*

Rheological measurements of emulsions were performed in a rotational rheometer (Haake Mars II, Thermo Scientific Corporation, Germany), equipped with a stainless steel coneplate geometry sensor (cone angle =1°; diameter = 60 mm; gap =0.052 mm) and maintained at 25.0 ± 0.1 °C by an ultrathermostatic bath (Phoenix 2C30P, Thermo Scientific Corporation, Germany). These measurements were carried out immediately after obtaining the emulsions. Flow curves (1st up cycle, down cycle, and 2nd up cycle) were obtained by continuously varying the shear rate from 0.1 to 300 s$^{-1}$ in 180 s. Newtonian model (Eq. (1)) and Power Law (Ostwald-de-Waele) model (Eq. (2)) [19] were tested to mathematically model experimental data.

$$\tau = \eta \cdot \dot{\gamma} \quad (1)$$

$$\tau = K \cdot (\dot{\gamma})^n \quad (2)$$

In Eqs. (1) and (2), $\tau$ is the shear stress, $\eta$ is the viscosity, $K$ is the consistency index, $n$ is the flow behavior index, and $\dot{\gamma}$ is the shear rate.

For dynamic oscillatory assays, the linear viscoelastic range of the emulsions was determined by initially performing a strain sweep (0.01 to 10%) at a constant frequency of 1 Hz. After that, the frequency sweeps were carried out, from 0.01 to 1 Hz. The constant strain amplitude was controlled in 1.0%, according to the previously determined linear viscoelastic range. The values of storage modulus ($G'$) and loss modulus ($G''$) as a function of the frequency ($\omega$) were continuously recorded. Size and $\zeta$-potential of Oil Dispersed Droplets Average sizes, size distributions, and $\zeta$-potential of the oil droplets of emulsions were evaluated by dynamic light scattering (DLS) (Zetasizer Nano-ZS, Malvern Instruments, United Kingdom), as described in more details elsewhere [15]. Briefly, emulsions were sampled at 0.5–1.0 cm below the surface and then diluted (1:250) to prevent multiple scattering effects, using as diluent the same chitosan dispersion or acid lactic solution used as a continuous phase in each case. The intensity of scattered light by the emulsions was much higher in comparison to the intensity of their diluents (Fig. 2). Therefore, interferences or additive effects of the diluent were not considered to droplet size and ζ-potential results. Size distributions were obtained by means of the amplitude of the decay rate, which is obtained by fitting the normalized temporal intensity correlation functions, by Non-Negative Least Square algorithm (NNLS) [20]. The electrophoretic mobility of the particles was calculated by measuring the average speed and the direction of oil droplets movement due to a controlled electric field and, then, the ζ-potential was calculated by assuming the Smoluchowski model for the double electrical layer [21].

*Macroscopic Stability*

Emulsion photographs were taken after preparation of emulsions and after a seven-day storage period. The extension of oil gravitational separation was quantified in terms of creaming index (CI) [14, 22] (Eq. (3)).

$$\mathrm{CI} = \frac{V_C}{V_T} \cdot 100\% \tag{3}$$

In Eq. (3), $V_T$ represents the total emulsion volume in measuring cylinders and $V_C$ is the creamed oil volume, measured after seven days.

**Interfacial Tension Measurements**

Chitosan [0.025; 0.050; and 0.100 g·(100 mL)$^{-1}$] was dispersed in aqueous lactic acid solution (pH 3.0), and stirred during 24 h, at room temperature. Tween 20 [0.1 mL·(100 mL)$^{-1}$] was added to these dispersions, and the interfacial tension between sunflower oil and these aqueous systems was measured using a pendant drop tensiometer (DAS-100, Krüss GmbH, Germany). Simultaneously, a control system without chitosan but with Tween 20 [0.1 mL·(100 mL)$^{-1}$] was prepared in deionized water. Another control system without chitosan and/or Tween 20 was prepared in aqueous lactic acid solution (pH 3.0). In each assay, a drop (10.0 ± 1.0 μL) of aqueous dispersion was formed in the bucket containing sunflower oil at 25.0 ± 1.0 °C. Measurements were performed each 30 s, during 900 s. The equilibrium interfacial tension ($\sigma_{eq}$) values were estimated using a long-time extrapolation of an empirical exponential decay model adjusted for σ = f(t) (Eq. 4).

$$\sigma = \sigma_{eq} + A \cdot e^{-b\sqrt{t}} \tag{4}$$

In Eq. (4), σ is the interfacial tension, $\sigma_{eq}$ is the equilibrium interfacial tension, $A$ is a fitting constant, $b$ is a constant related to decay rate of σ until reaching the equilibrium value $\sigma_{eq}$, and $t$ is the time.

**Data Analyses**

All measurements were carried out in three repetitions and results were reported as mean ± standard deviation. Data were submitted to analysis of variance (ANOVA) ($p < 0.05$) or paired t test ($p < 0.05$), using the SAS software (version 9.3, SAS Institute Incorporation, USA), licensed by the Universidade Federal de Viçosa. When pertinent, means were compared through Tukey's test ($p < 0.05$). The adequacy of fitting of all regression models was evaluated in terms of both coefficient of determination ($R^2$) and mean absolute percentage error (MAPE) (Eq. (5)).

$$\mathrm{MAPE} = \frac{1}{n}\sum_{i=1}^{n}\left|\frac{(Y_i - \hat{Y}_i)}{Y_i}\right| \cdot 100\% \tag{5}$$

In Eq. (5), $Y_i$ is the i$^{th}$ experimental score, $\hat{Y}_i$ is the i$^{th}$ score predicted applying the adjusted model and $n$ is the number of predicted/experimental score pairs. To be considered adequately fitted to experimental data, models had to present $R^2$ values ≥0.9 and MAPE values ≤10%.

**Results**

**Rheological Characterization of Emulsions**

None of the emulsions showed hysteresis in their rheogram, as the three flow curves (1st up cycle, down cycle, and 2nd up cycle) were superposed. Therefore, only the 2nd up cycle curves were represented in rheograms (Supplementary Material). As the emulsions did not present critical stress for flowing (i.e., all of them had $\tau_0 = 0$), Newtonian and Power Law models were tested to mathematically describe their flow behavior. Even though both models fitted adequately ($R^2 \geq 0.98$ and MAPE ≤ 6.2%) to experimental data for $\tau = f(\dot{\gamma})$, one can easily perceive the linearity of the flow curves, which point out the unsuitability of the power law model for representing in practice the rheological behavior of these emulsions. Therefore, results indicated that these emulsions can be considered Newtonian fluids. As presented in Table 1, the viscosities values ($\eta$) of emulsions containing chitosan (Q30, Q35, and Q40) did not show significant differences among themselves ($p > 0.05$). The same was observed when comparing the counterparts without chitosan (A30, A35, and A40). However, a significant difference ($p < 0.05$) in terms of $\eta$ was observed between emulsions with and without chitosan, at a given pH.

Dynamic oscillatory assays (frequency sweeps from 0.1 to 1 Hz) were also performed. Results are shown in Fig. 1. For Q30 and Q35, $G''$ values increased from 0.4 to 0.6 Pa with the increase of the shear frequency, whereas $G'$ values

**Table 1** Viscosity values ($\eta$) of the emulsions and adequacy of fitting of the Newtonian model in modeling their flow behavior.

| Emulsion | $\eta$ (mPa·s) | Adequacy of fitting | |
|---|---|---|---|
| | | $R^2$ | MAPE (%) |
| Q30 | $2.849 \pm 0.004^a$ | 0.99 | 1.46 |
| A30 | $1.218 \pm 0.001^b$ | 0.99 | 0.90 |
| Q35 | $2.978 \pm 0.010^a$ | 0.99 | 4.74 |
| A35 | $1.223 \pm 0.003^b$ | 0.99 | 3.06 |
| Q40 | $2.824 \pm 0.010^a$ | 0.99 | 3.25 |
| A40 | $1.209 \pm 0.006^b$ | 0.99 | 6.22 |

Different letters in the same column represent significant differences between emulsions, by Tukey test ($p < 0.05$).

increased more sharply from 0.6 to 1.1 Pa within the same shear frequency range. On the other hand, for Q40, both $G'$ and $G''$ values presented less expressive augmentation, $G''$ from 0.2 to 0.4 Pa and $G'$ from 0.3 to 0.6 Pa, as the shear frequency was increased. As $G' > G''$, all of these emulsions containing chitosan presented a predominantly elastic character, especially under mechanical solicitation. Emulsions without chitosan (A30, A35, and A40) presented a different trend: $G'$ and $G''$ values were very similar within the studied frequency ranges (about 0.35 Pa), indicating no predominance of neither elastic nor viscous character.

## Droplets Sizes

DLS results are compiled in Table 2. Emulsions without chitosan had unimodal droplet size distributions at $t = 0$. Average diameters measured at $t = 0$ were of 275, 390, and 408 nm, for A30, A35, and A40 emulsions, respectively. Differently, emulsions with chitosan had bimodal droplet size distributions at $t = 0$ (Fig. 2). The main population of droplet (that scatters around 90% of the total scattered intensity) presented diameters of 336–357 nm. No significant differences ($p > 0.05$) were found among the sizes of this population of larger droplets present in emulsions with chitosan and those of droplets of emulsions without chitosan. In addition, in emulsions containing chitosan, a fraction of droplets with diameters of a few dozen nanometers was detected (around 10% of the total scattered intensity). Their averages diameters were 31, 50, and 43 nm, respectively for Q30, Q35, and Q40 emulsions. A significant difference ($p < 0.05$) was observed between the sizes of this population of smaller droplet and those of the main population within the same emulsion. It is important to note that the relative abundance of the two populations is not equal to the ratio of the intensity from each population since the intensity increases dramatically with the size of the scatters ($\propto$ size$^6$, neglecting the dependence with the scattered angle). Also, it is expected that the intensity of each population changes with

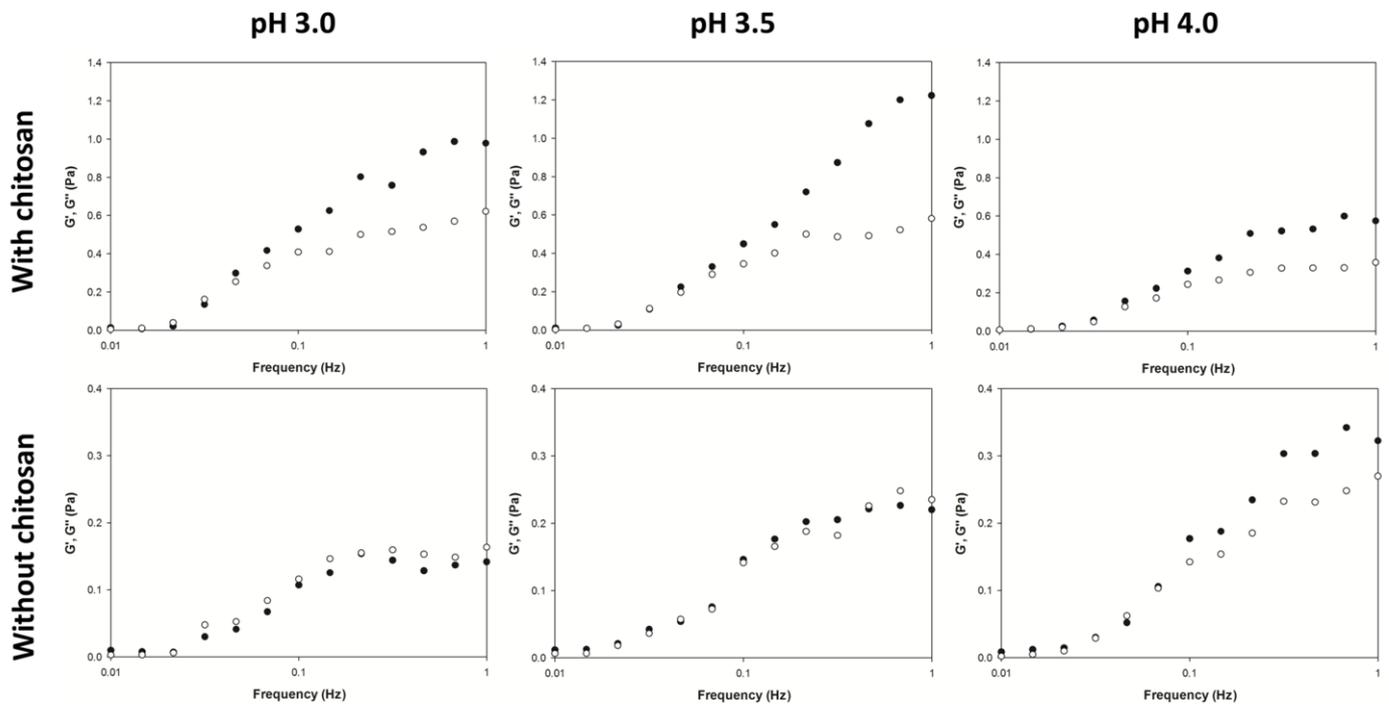

**Figure 1** $G'$ (●) and $G''$ (○) in frequency sweeps for emulsions with and without chitosan, at pH = 3.0, 3.5 and 4.0.

**Table 2** Surface electrical charges and average diameters ($d$) of emulsions' droplets, at $t = 0$ and $t = 7$ days.

| Emulsion | Droplets surfaces electrical charges | | Droplets sizes | | |
|---|---|---|---|---|---|
| | $\zeta$-potential (mV) at $t = 0$ | $\zeta$-potential (mV) at $t = 7$ days | Type of distribution | $d$ (nm) at $t = 0$ day | $d$ (nm) at $t = 7$ days |
| Q30 | 54 ± 1$^{a,A}$ | 53 ± 1$^{a,A}$ | Bimodal | 336 ± 14$^{a,A}$ | 364 ± 8$^{a,A}$ |
| | | | | 31 ± 1$^{b,A}$ | 45 ± 10$^{b,A}$ |
| A30 | -1 ± 2$^{b,A}$ | -1 ± 1$^{b,A}$ | Unimodal | 275 ± 85$^{a,A}$ | 297 ± 14$^{a,A}$ |
| Q35 | 51 ± 2$^{a,A}$ | 50 ± 2$^{a,A}$ | Bimodal | 357 ± 37$^{a,A}$ | 355 ± 14$^{a,A}$ |
| | | | | 50 ± 16$^{b,A}$ | 40 ± 15$^{b,A}$ |
| A35 | -2 ± 2$^{b,A}$ | -4 ± 3$^{b,A}$ | Unimodal | 390 ± 268$^{a,A}$ | 317 ± 37$^{a,A}$ |
| Q40 | 52 ± 2$^{a,A}$ | 51 ± 4$^{a,A}$ | Bimodal | 354 ± 37$^{a,A}$ | 372 ± 56$^{a,A}$ |
| | | | | 43 ± 11$^{b,A}$ | 56 ± 9$^{b,A}$ |
| A40 | -3 ± 2$^{b,A}$ | -5 ± 3$^{b,A}$ | Unimodal | 408 ± 135$^{a,A}$ | 304 ± 5$^{a,A}$ |

Different small letters in the same column represent significant differences between emulsions, by Tukey test ($p < 0.05$).
Different capital letters in the same line represent significant differences for an emulsion between $t = 0$ and $t = 7$ days, by t test ($p < 0.05$).

the scattered angle if $qd \gtrsim 1$ ($q$ is the modulus of the scattering vector and $d$ the size of the scatters). For the studied systems $q = 0.026$ nm$^{-1}$ leading to $d \gtrsim 38$ nm as the condition to have peaks that depends of the scattered angle.

When comparing average diameters values at $t = 0$ and $t = 7$ days for a given droplet population of each emulsion, no significant differences were detected in any case ($p > 0.05$). These results indicate that emulsification process parameters set were successful in forming emulsions with average droplet sizes stable in practice for at least 7 days. Additionally, chitosan added to the continuous phase of the emulsion conferred to them distinct droplet size distributions, which can be related to their rheological differences (more pronounced elastic character) regarding the corresponding emulsions without chitosan at the same pH value.

**Droplets $\zeta$-Potential**

$\zeta$-potential results are also shown in Table 2. Emulsions without chitosan presented $\zeta$-potential values slightly negative but near zero. These results were not surprising due to the electrostatic neutrality of Tween 20 molecules adsorbed on the droplet surface. The slightly negative value of $\zeta$-potential could be due to free fatty acids present in the dispersed phase [23, 24] or lactate anions interacting with Tween 20 molecules on the O/W interface. Conversely, emulsions formulated with chitosan exhibited strongly positive $\zeta$-potential values ~50 mV, and significant differences were observed ($p < 0.05$) between emulsions with or without chitosan. However, no significant changes ($p > 0.05$) were observed in $\zeta$-potential of the droplets among emulsions containing chitosan but with different pH values. Also, no significant changes ($p > 0.05$) were observed in $\zeta$-potential of the droplets of the emulsions at $t = 0$ and $t = 7$ days.

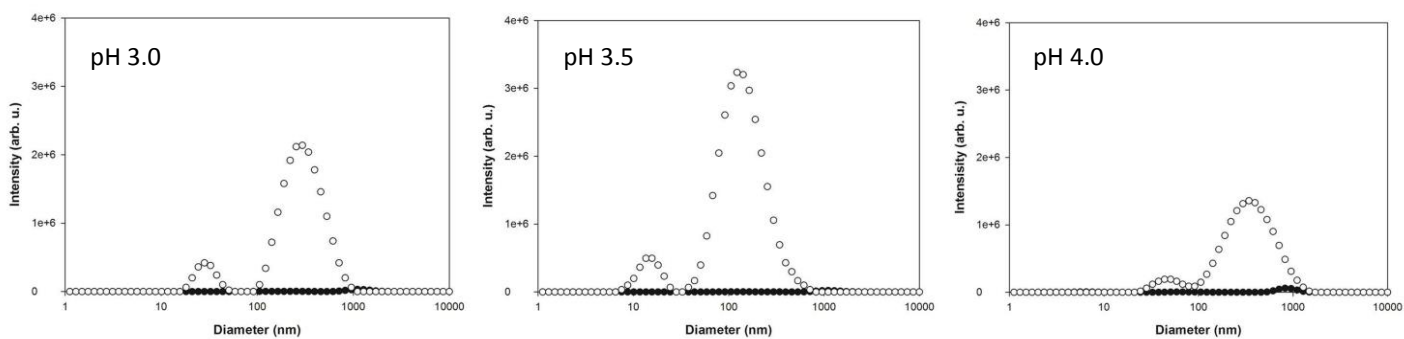

**Figure 2** - Intensity of light scattering and diameter distribution of droplets in emulsions containing chitosan [0.1 g·(100 mL)$^{-1}$], with pH = 3.0, 3.5 and 4.0, represented by empty circles (○). In each case, black circles (●) represent the corresponding data for diluents used for performing DLS analyses (i.e., the aqueous dispersion used to form the continuous phases of each emulsion, i.e., Tween 20 [1 mL·(100 mL)$^{-1}$] + chitosan [0.1 g·(100 mL)$^{-1}$] in lactic acid solutions at the same pH). The curves were normalized to have the same incident intensity, showing that the signals of the diluents were insignificant compared to those of the emulsions.

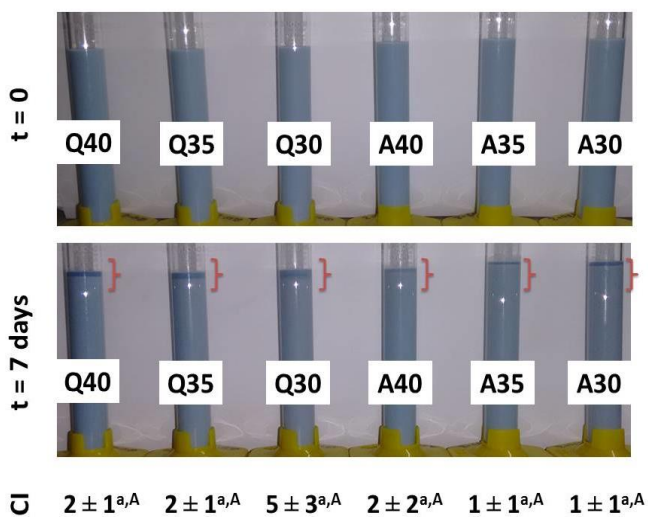

**Figure 3** - Photographs of emulsions with or without chitosan in the continuous phase with different pH values, at $t = 0$ and $t = 7$ days. Different small letters represent significantly different cream indexes values (CI; %) between emulsions, by ANOVA ($p < 0.05$).

**Macroscopic Stability**

Emulsions were photographed after their production ($t = 0$) and after a seven day storage period ($t = 7$) (Fig. 3). The bluish, opaque tonality of the emulsions is due to the hydrophobic dye Sudam Black B previously added to the sunflower oil. Evident signs of creaming were not observed in any of the emulsion at $t = 0$. However, all emulsions had on top of the test tube a very thin layer with a blue tonality slightly more translucent, at $t = 7$. This suggests that creaming phenomena were likely to be starting after this storage time. Based on these layers thickness, cream index values were calculated as $\sim 1-2\%$. No significant difference in creaming was observed ($p > 0.05$) between emulsions with different pH, neither with regard to the presence or absence of chitosan in the continuous phase.

**Interfacial Tension**

Interfacial tension versus time curves for O/W systems containing 0.1 mL·(100 mL)$^{-1}$ Tween 20 and chitosan [0.025, 0.050, or 0.100 g·(100 mL)$^{-1}$] in aqueous solution lactic acid (pH 3.0) showed an exponential decrease of the interfacial tension, as did the control system containing 0.1 mL·(100 mL)$^{-1}$ Tween 20 in the aqueous phase, but without chitosan. All these curves showed a very similar temporal decreasing profile (Fig. 4). An empirical exponential model (Eq. (4)) was adjusted to these experimental data and the corresponding parameters are given in Table 3. This empirical model fitted well to experimental for $\sigma = f(t)$ data ($R^2 \geq 0.96$ and MAPE $\leq 1.1\%$). Systems containing chitosan showed $\sigma_{eq}$ varying from 56.40 to 57.85 mN·m$^{-1}$, $A$ from 21.74 to 25.37 mN·m$^{-1}$, and $b = 0.08$ s$^{-0.5}$. Systems without chitosan presented $\sigma_{eq}$ from 46.77 to 54.80 mN·m$^{-1}$, $A$ from 17.07 to 24.85 mN·m$^{-1}$, and $b$ from 0.03 to 0.06 s$^{-0.5}$. However, no statistically significant difference ($p > 0.05$) for $\sigma_{eq}$, $A$, and $b$ values was found among these systems. The model could not be adjusted to experimental data of the control which contained only lactic acid in the aqueous phase. Not surprisingly, the interfacial tension in this case remained practically constant ($\sim 70$ mN·m$^{-1}$) and higher than $\sigma_{eq}$ obtained for the other systems, since lactic acid molecules are not amphiphilic.

**Discussion**

Deacetylation degree (*DD*) and viscometric molar mass ($M_v$) influence various physicochemical properties of chitosan acidic dispersions [1]. $DD > 75\%$ have been correlated to an easier dispersibility of chitosan in acidic aqueous media [15, 25]. It was corroborated by our results, as chitosan with $DD \sim 85\%$ formed aqueous dispersions in which the biopolymer concentration [0.1 g·(100 g)$^{-1}$] was compatible with those usually employed for other polysaccharides with thickening/ stabilizing purposes in formulated food products.

For O/W emulsions elaborated using such aqueous dispersions, with pH ranged between 3.0 and 4.0, significant differences ($p < 0.05$) were not observed among their viscosities. Indeed, as chitosan amino groups present pKa $\sim 6.4$ [26], such groups are predominantly protonated (thus positively charged) at pH $\leq 4.0$. Therefore, in addition to repulsive electrostatic interactions between chitosan chains, attractive chitosan-water interactions (e.g. ion-dipole interactions) are expected to be favored, which can macroscopically trigger a viscosity increasing. No significant visual difference was observed among emulsions with different pH, indicating that different lactate concentrations within the systems were not capable to promote noticeable visual changes in these emulsions. Instead, when comparing emulsions with and without chitosan, at a given pH, significant differences ($p < 0.05$) were found among their viscosity values. Emulsions containing 0.1 g·(100 g)$^{-1}$ of chitosan presented viscosities about 3 times higher than those of their counterparts without chitosan. In other words, the addition of chitosan, even at a low concentration, had a significant thickening effect. Other polysaccharides commonly used as thickeners in industrial applications, such as the electrically neutral guar and locust bean gums, usually require at least the double of this concentration to achieve a similar thickening effect [27].

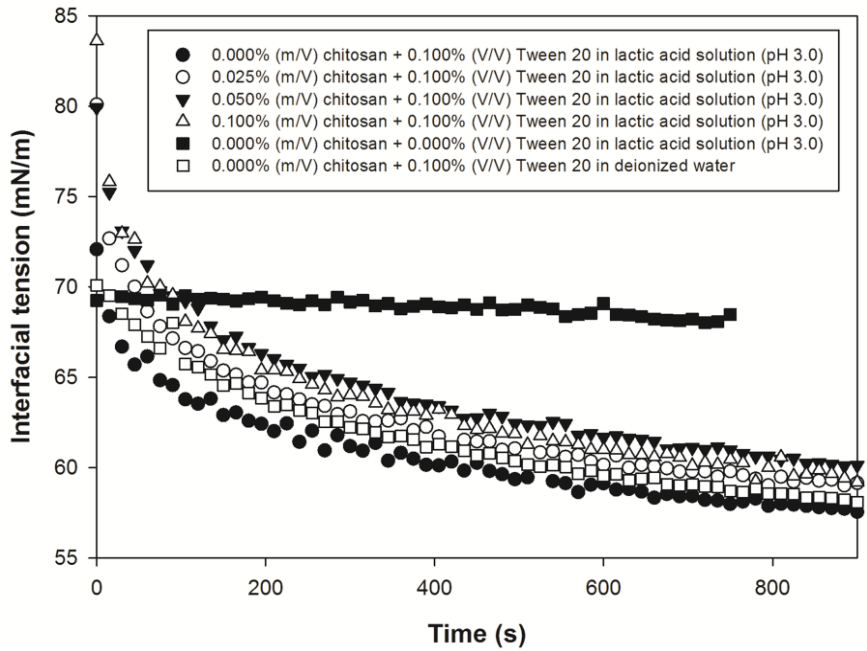

**Figure 4** – Decay profiles of interfacial tension between sunflower oil in contact with different aqueous phases containing chitosan and/or lactic acid and/or Tween20.

In dynamic oscillatory assays, emulsions containing chitosan presented values of both $G''$ and $G'$ higher than those without chitosan in rheological dynamic oscillatory assays (frequency sweeps at constant temperature of 25 °C). Furthermore, systems with chitosan presented $G' > G''$. These results indicate that such emulsions had a predominantly elastic character and can suggest that adding chitosan contributes to a physical structuration of the medium in some extent. In the present case, this elastic character can be related to a reduced mobility of the oil droplets, thus contributing to improve the kinetic stability of the emulsions. Calero et al. [12] analyzed emulsions formulated with potato protein (2 g·(100 mL)$^{-1}$) and chitosan (0.25–1.00 g·(100 mL)$^{-1}$), and also reported $G' > G''$ (at 1 rad·s$^{-1}$) for these preparations. After a 15- day storage period, additional

**Table 3** Model parameters and adequacy of fitting for the empirical exponential model adjusted to data of interfacial tension decay between sunflower oil in contact with different aqueous phases containing chitosan and/or lactic acid and/or Tween20, at pH = 3.0.

| Aqueous systems | Adjusted parameters for $\sigma = \sigma_{eq} + A \cdot e^{-b\sqrt{t}}$ (Eq.4) | | | Adequacy of fitting | |
|---|---|---|---|---|---|
| | $\sigma_{eq}$ (mN·m$^{-1}$) | $A$ (mN·m$^{-1}$) | $b$ (s$^{-0.5}$) | $R^2$ | MAPE (%) |
| 0.000 g·100 mL$^{-1}$ chitosan + 0.0 mL·(100 mL)$^{-1}$ Tween 20 in LA | - | - | - | - | - |
| 0.000 g·100 mL$^{-1}$ chitosan + 0.1 mL·(100 mL)$^{-1}$ Tween 20 in LA | 54.80 ± 6.15$^a$ | 17.07 ± 2.51$^a$ | 0.06 ± 0.02$^a$ | 0.99 | 0.40 |
| 0.025 g·100 mL$^{-1}$ chitosan + 0.1 mL·(100 mL)$^{-1}$ Tween 20 in LA | 57.85 ± 2.52$^a$ | 21.74 ± 1.00$^a$ | 0,08 ± 0.01$^a$ | 0.96 | 1.00 |
| 0.050 g·100 mL$^{-1}$ chitosan + 0.1 mL·(100 mL)$^{-1}$ Tween 20 in LA | 56.40 ± 2.36$^a$ | 23.47 ± 1.17$^a$ | 0.08 ± 0.01$^a$ | 0.99 | 0.25 |
| 0.100 g·100 mL$^{-1}$ chitosan+ 0.1 mL·(100 mL)$^{-1}$ Tween 20 in LA | 57.75 ± 0.51$^a$ | 25.37 ± 0.72$^a$ | 0.08 ± 0.03$^a$ | 0.99 | 0.61 |
| 0.000 g·100 mL$^{-1}$ chitosan + 0.1 mL·(100 mL)$^{-1}$ Tween 20 in water | 46.77 ± 1.97$^a$ | 24.85 ± 1.78$^a$ | 0.03 ± 0.00$^a$ | 0.98 | 1.08 |

Different letters in the same column represent significant differences between the adjusted values of the model parameter (Eq. 4), by ANOVA ($p < 0.05$).
LA: lactic acid solution.

analyses showed that values of both $G'$ and $G''$ increased compared to those previously measured, suggesting some time-dependent structuration in the systems. Nonetheless, comparisons should be done with cautiousness, mainly because of two factors: firstly, Calero et al. [12] worked with chitosan concentrations much higher than that used in the present study. Secondly, proteins were used as emulsifier in their study, whilst here a small molecular mass, electrically neutral surfactant (Tween 20) was used. It is well known that the net electrical charge of proteins is dependent on the pH of the medium, and that protein and polysaccharides may establish numerous specific interactions which may play a determinant role in the thermodynamic and mechanical behaviors of emulsion systems.

Anyhow, the increase of the elastic character attributed to chitosan in the emulsions contributes to reduce the Brownian motion of emulsion droplets, as well as the frequency and intensity of collisions among them. Therefore, from this point of view, chitosan is expected to enhance the kinetic stability of emulsions, which may have physical implications either macroscopically or microscopically detectable. Macroscopically, cream index values measured for emulsions with and without chitosan, at each studied pH, had no significant difference ($p > 0.05$) among them, after a 7-day storage period at 25 °C. When analyzed at a microscopic level, emulsions containing chitosan presented considerable differences in terms of droplet average size and size distribution, in relation to the counterparts without chitosan, as pointed out by dynamic light scattering results.

Soon after their fabrication ($t = 0$), emulsions without chitosan presented unimodal droplet size distributions, with average diameters ~ 297–318 nm. Differently, emulsions formulated with chitosan presented bimodal droplet size distributions: a first population of droplets with average diameters 40–57 nm and a second population with average diameters 355–373 nm. In other words, in emulsions whose continuous phases contained chitosan previously dispersed, there were a fraction of droplets which were about 7–8 times smaller than the remaining. A reasonable explanation for this finding could be the fact that, in the very few seconds after the homogenization operation, emulsions both with and without chitosan were likely to present a more uniform droplet size distribution, with a smaller average size. In those without chitosan, the coalescence phenomenon [28] starts earlier and occurs at a higher rate, generating bigger droplets more quickly. Conversely, in emulsions with chitosan, as the biopolymer reduces the coalescence rate, a fraction of the smallest droplets initially formed remained longer in the medium, and they could be detected few minutes later in DLS analyses. Kaasgaard & Keller [13], Klinkesorn & Namatsila [14], and Mun et al. [24], which also studied other emulsified systems containing chitosan, reported unimodal size distributions of droplets sizes either in emulsions using chitosan or in controls without it. Nevertheless, chitosans used by these authors had different $DD$, average molecular masses and came from different furnishers (biopolymer concentrations and type of acid were also different in each case). This difficult in making comparisons emphasizes the importance of characterizing each chitosan sample to be used in a study, at least in terms of molecular mass and $DD$ and $M_v$, prior to any other assays. After storage for 7 days, no significant differences in average sizes of disperse phase particles were detected ($p > 0.05$), compared to DLS results obtained at $t = 0$. In fact, in some cases, depletion flocculation may occur in emulsion systems containing non-adsorbing polysaccharide in the continuous phase, above a certain concentration [29, 30]. At this concentration, when two adjacent droplets approach each other, the space between them is devoid of polysaccharide, which drives an osmotic gradient to remove the solvent in this space, causing flocculation. If this phenomenon had occurred in the emulsions containing chitosan, an augmentation of average particles sizes, or even a more drastic visually detectable flocculation, would have been expected. The biopolymer concentration of 0.1 g·(100 g)$^{-1}$ was effective in change the rheological characteristics of the emulsions without triggering depletion flocculation during storage for 7 days at 25 °C.

Concerning the electrical charges on the droplets surfaces, oil droplets of the emulsions with chitosan showed highly positive ζ-potential values (~ +50 mV), whereas those of emulsions without chitosan presented ζ-potential slightly negative, but near zero ($|ζ| ≤ 5$ mV). These values and signs were stable during 7 days ($p > 0.05$). Tween 20 molecules on the droplet O/W interfaces probably interacted with lactate anions, which would explain these slight negative charges in systems without chitosan. With chitosan, however, the strongly positively charged biopolymer chains integrate the electrical double layer of droplets, as observed for similar chitosan emulsions studied by Kaasgaard & Keller [13], Klinkesorn & Namatsila [14], and Mun et al. [24], when working with organic other acids and emulsifiers (CITREM LR10, Tween 80 and Tween 20, respectively). The increase of $|ζ|$ is frequently related to an improved kinetic stability of emulsions, since it increases electrostatic repulsion among oil droplets [31]. Besides this electrostatic effect, the chitosan chains around the droplets may provide steric hindrances among them, further enhancing the kinetic stability of their sizes along the studied period of time.

Finally, differences were not observed ($p > 0.05$) in $σ_{eq}$ or in interfacial tension decay rates (related to the b parameter in Eq. 4) of systems with or without chitosan. Then, no direct interfacial/emulsifying effect can be attributed to chitosan, although a conjoint analysis of the results indicates that this biopolymer, when previously dispersed in aqueous solutions of lactic acid, contributes to improve the kinetic stability of O/W emulsions through different possible mechanisms.

## Conclusions

Emulsions containing chitosan previously dispersed in lactic acid aqueous solutions were successfully obtained. Therefore, acetic acid, which is typically used to disperse chitosan in aqueous media, can be replaced by another food-grade organic acid, allowing the production of emulsified formulations without the frequently undesirable acetic acid flavor. Chitosan was shown an effective thickener agent in these systems even at only 0.1 g·(100 g)$^{-1}$, as pointed by two different rheological assays. In addition, the presence of chitosan modified both the size distributions and average sizes of droplets in emulsions. The surfaces of oil droplets in emulsions containing chitosan showed highly positive ζ-potential values. This point is a finding of particular technological importance, as the electrostatic repulsion among droplets is one of the major factors enhancing the kinetic stability of emulsions. Neither the average diameters nor the ζ-potential values presented significant changes after storage of the emulsions at 25 °C during 7 days. The overall results of this study corroborates the hypothesis that using low concentration of chitosan dispersed in lactic acid aqueous solutions is a strategic alternative to be considered as starting point when formulating different new emulsion-based food products (e.g.: dairy desserts and some sauces), combining biofuncional claims and thickener/ stabilizer properties. Additional investigations are now in progress, in order to precisely understand the structuration of the chemical species around the oil droplets in these emulsions. In fact, at this point, the results did not allow us concluding whether chitosan chains found their way at the interface (in a mixed layer), or close to it (as a second layer), or still dispersed at the bulk.

## Aknowledgements

The authors are very grateful to: Brazilian research agencies CNPq (scholarship of L. S. Soares), FAPEMIG and FUNARBE-UFV (financial support); Prof. Luciano de Moura Guimarães (DPF-UFV) and Prof. Cristiano Fantini Leite (UFMG) for their precious help in Raman spectroscopy analyses; Dr. Eber Antonio Alves Medeiros (DTA-UFV), Prof. Nilda de Fátima Ferreira Soares (DTA-UFV) and Prof. Nélio José de Andrade (DTA-UFV), which kindly allowed us using the instrumental apparatus needed to perform ζ-potential and interfacial tension analyses; Prof. Luis Henrique Mendes da Silva (DEQ-UFV), for the rewarding discussions.

# Rheological and physicochemical studies on emulsions formulated with chitosan previously dispersed in aqueous solutions of lactic acid

Lucas de Souza Soares[a], Janaína Teles de Faria[b], Matheus Lopes Amorim[a],

João Marcos de Araújo[c], Luis Antonio Minim[a], Jane Sélia dos Reis Coimbra[a],

Alvaro Vianna Novaes de Carvalho Teixeira[d], Eduardo Basílio de Oliveira[a] ✉

[a] *Departamento de Tecnologia de Alimentos (DTA), Universidade Federal de Viçosa (UFV), Campus Universitário, CEP 36570-900, Viçosa, MG, Brazil.*

[b] *Instituto de Ciências Agrárias (ICA), Universidade Federal de Minas Gerais (UFMG), CEP 39400-000, Montes Claros, MG, Brazil.*

[c] *Departamento de Biologia Geral (DBG), Universidade Federal de Viçosa (UFV), Campus Universitário, CEP 36570-900, Viçosa, MG, Brazil.*

[d] *Departamento de Física (DPF), Universidade Federal de Viçosa (UFV), Campus Universitário, CEP 36570-900, Viçosa, MG, Brazil.*

✉ Corresponding author: eduardo.basilio@ufv.br

I. **Viscosities and refraction indexes of dispersions used in DLS analyses**

Table SM1 - Physical properties of acid dispersions (AD) in different pH (3.0, 3.5 and 4.0) values with 1 mL·(100 mL)$^{-1}$ Tween 20 and 0.1 g·(100 mL)$^{-1}$ chitosan

| Acid dispersion | Index of refraction | Viscosity (mPa·s) |
|---|---|---|
| AD-Q40 | 1.3331 ± 0.0002 | 4.05 ± 0.20 |
| AD-Q35 | 1.3332 ± 0.0001 | 4.05 ± 0.30 |
| AD-Q30 | 1.3333 ± 0.0001 | 3.72 ± 0.20 |

II. **Raman spectroscopy results**

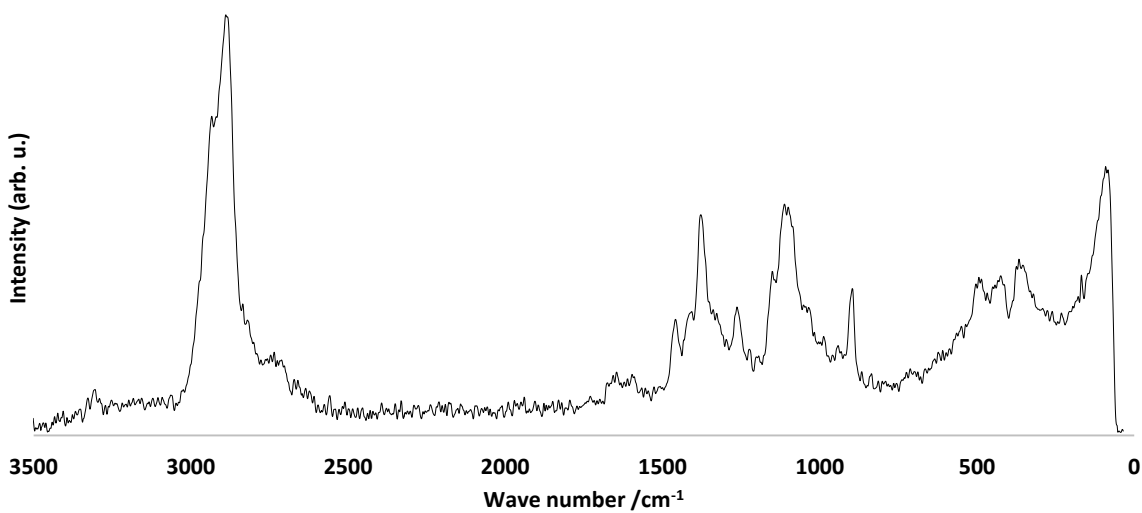

Figure SM1 - Raman spectrum of chitosan used in the present study

## III. Rheograms of the emulsions

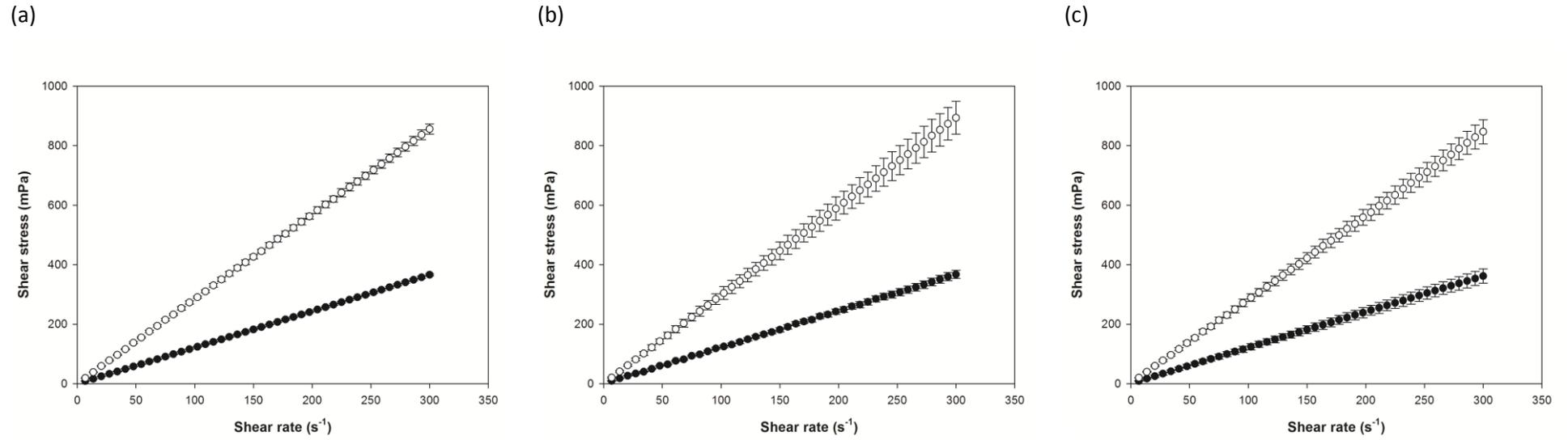

Figure SM2 - Flow curves of (○) Q30 and (●) A30 (a), (○) Q35 and (●) A35 (b), and (○) Q40 and (●) A40 (c) emulsions

## IV. Statistical treatment of rheograms (Figure SM2)

Table SM2 - Nonlinear OLS Summary of Residual Errors to Newtonian Model adjusted to experimental data

|     | DF Model | DF Error | MSE     | Adjusted R-Square |
|-----|----------|----------|---------|-------------------|
| Q30 | 1        | 131      | 60.6969 | 0.9990            |
| A30 | 1        | 131      | 2.8005  | 0.9998            |
| Q35 | 1        | 131      | 760.4   | 0.9888            |
| A35 | 1        | 131      | 44.7217 | 0.9960            |
| Q40 | 1        | 131      | 358.8   | 0.9941            |
| A40 | 1        | 131      | 149.5   | 0.9865            |

Table SM3 - Nonlinear OLS Parameter Estimates to Newtonian Model adjusted to experimental data

|     | Parameter | Approx. Std. Error | t Value | Approx. Pr > \|t\| |
|-----|-----------|--------------------|---------|--------------------|
| Q30 | $\mu$     | 0.00385            | 740.17  | < 0.0001           |
| A30 | $\mu$     | 0.000827           | 1473.42 | < 0.0001           |
| Q35 | $\mu$     | 0.0136             | 218.54  | < 0.0001           |
| A35 | $\mu$     | 0.00330            | 370.19  | < 0.0001           |
| Q40 | $\mu$     | 0.0135             | 209,72  | < 0.0001           |
| A40 | $\mu$     | 0.00604            | 200.03  | < 0.0001           |

Table SM4 - Nonlinear OLS Summary of Residual Errors to Power Law Model adjusted to experimental data

|     | DF Model | DF Error | MSE     | Adjusted R-Square |
|-----|----------|----------|---------|-------------------|
| Q30 | 2        | 130      | 61.1363 | 0.9990            |
| A30 | 2        | 130      | 2.7644  | 0.9998            |
| Q35 | 2        | 130      | 766.2   | 0.9887            |
| A35 | 2        | 130      | 45.0624 | 0.9960            |
| Q40 | 2        | 130      | 364.4   | 0.9940            |
| A40 | 2        | 130      | 150.4   | 0.9864            |

Table SM5 - Nonlinear OLS Parameter Estimates to Power Law Model adjusted to experimental data

|     | Parameter | Approx. Std. Error | t Value | Approx. Pr > \|t\| |
|-----|-----------|--------------------|---------|--------------------|
| Q30 | $K$       | 0.0628             | 45.61   | < 0.0001           |
|     | $n$       | 0.00407            | 245.70  | < 0.0001           |
| A30 | $K$       | 0.0131             | 91.45   | < 0.0001           |
|     | $n$       | 0.00203            | 494.78  | < 0.0001           |
| Q35 | $K$       | 0.2209             | 13.45   | < 0.0001           |
|     | $n$       | 0.0138             | 72.56   | < 0.0001           |
| A35 | $K$       | 0.0535             | 22.78   | < 0.0001           |
|     | $n$       | 0.00814            | 122.93  | < 0.0001           |
| Q40 | $K$       | 0.2209             | 12.90   | < 0.0001           |
|     | $n$       | 0.0144             | 69.30   | < 0.0001           |
| A40 | $K$       | 0.1017             | 12.40   | < 0.0001           |
|     | $n$       | 0.0150             | 66.30   | < 0.0001           |

Table SM6 - ANOVA used to evaluate $K$ and $n$ parameter from Power Law Model

| Parameter | DF | Mean Square | F Value | Pr > F   |
|-----------|----|-------------|---------|----------|
| $K$       | 5  | 2.49732970  | 164.05  | < 0.0001 |
| $n$       | 5  | 0.00003653  | 0.51    | 0.7657   |

## V. Statistical treatment of data presented in Figure 2

Table SM7 - ANOVA used to evaluate $d_h$ and ζ-potential of the emulsions

|  | DF | Mean Square | F Value | Pr > F |
|---|---|---|---|---|
| $d$ ($t$ = 0) | 8 | 69852.2859 | 18.03 | < 0.0001 |
| $d$ ($t$ = 7) | 8 | 64305.7152 | 71.86 | < 0.0001 |
| ζ potential ($t$ = 0) | 5 | 2631.31987 | 726.46 | < 0.0001 |
| Z potential ($t$ = 7) | 5 | 2737.05594 | 433.43 | < 0.0001 |

Table SM8 - t test used to evaluate $d$ of the emulsions between $t$ = 0 and $t$ = 7 days

|  | Method | Variances | DF | t Value | Pr > \|t\| |
|---|---|---|---|---|---|
| Q30-1 | Pooled | Equal | 4 | -3.07 | 0.0571 |
|  | Satterthwaite | Unequal | 3.32 | -3.07 | 0.0576 |
| Q30-2 | Pooled | Equal | 4 | -2.25 | 0.0875 |
|  | Satterthwaite | Unequal | 2.03 | -2.25 | 0.1515 |
| A30 | Pooled | Equal | 4 | -1.26 | 0.2778 |
|  | Satterthwaite | Unequal | 2.95 | -1.26 | 0.2997 |
| Q35-1 | Pooled | Equal | 4 | 0.03 | 0.9768 |
|  | Satterthwaite | Unequal | 2.3 | 0.03 | 0.9778 |
| Q35-2 | Pooled | Equal | 4 | 0.82 | 0.4569 |
|  | Satterthwaite | Unequal | 3.98 | 0.82 | 0.4571 |
| A35 | Pooled | Equal | 4 | -0.41 | 0.7040 |
|  | Satterthwaite | Unequal | 2.48 | -0.41 | 0.7157 |
| Q40-1 | Pooled | Equal | 4 | -0.45 | 0.6738 |
|  | Satterthwaite | Unequal | 3.46 | -0.45 | 0.6772 |
| Q40-2 | Pooled | Equal | 4 | -1.57 | 0.1906 |
|  | Satterthwaite | Unequal | 3.93 | -1.57 | 0.1920 |
| A40 | Pooled | Equal | 4 | 1.25 | 0.2791 |
|  | Satterthwaite | Unequal | 2.58 | 1.25 | 0.3123 |

Table SM9 - t test used to evaluate ζ potential of the emulsions between $t$ = 0 and $t$ = 7 days

|  | Method | Variances | DF | t Value | Pr > \|t\| |
|---|---|---|---|---|---|
| Q30 | Pooled | Equal | 4 | 1.53 | 0.2001 |
|  | Satterthwaite | Unequal | 3.9 | 1.53 | 0.2018 |
| A30 | Pooled | Equal | 4 | -0.02 | 0.9876 |
|  | Satterthwaite | Unequal | 2.29 | -0.02 | 0.9881 |
| Q35 | Pooled | Equal | 4 | 0.45 | 0.6744 |
|  | Satterthwaite | Unequal | 3.52 | 0.45 | 0.6774 |
| A35 | Pooled | Equal | 4 | 0.80 | 0.4700 |
|  | Satterthwaite | Unequal | 2.9 | 0.80 | 0.4854 |
| Q40 | Pooled | Equal | 4 | 0.46 | 0.6668 |
|  | Satterthwaite | Unequal | 3.53 | 0.46 | 0.6698 |
| A40 | Pooled | Equal | 4 | 1.91 | 0.1295 |
|  | Satterthwaite | Unequal | 3.31 | 1.91 | 0.1442 |

## VI. Statistical treatment of data presented in Figure 3

Table SM10 - ANOVA used to evaluate cream index (CI)

|    | DF | Mean Square | F Value | Pr > F |
|----|----|-------------|---------|--------|
| CI | 5  | 5.55555556  | 2.22    | 0.1194 |

Table SM11 - t test used to evaluate CI of the emulsions between $t = 0$ and $t = 7$ days

|     | Method       | Variances | DF | t Value | Pr > \|t\| |
|-----|--------------|-----------|----|---------|------------|
| Q30 | Pooled       | Equal     | 4  | -2.80   | 0.0488     |
|     | Satterthwaite| Unequal   | 2  | -2.80   | 0.1074     |
| A30 | Pooled       | Equal     | 4  | -1.00   | 0.3739     |
|     | Satterthwaite| Unequal   | 2  | -1.00   | 0.4226     |
| Q35 | Pooled       | Equal     | 4  | -3.46   | 0.0527     |
|     | Satterthwaite| Unequal   | 2  | -3.46   | 0.0742     |
| A35 | Pooled       | Equal     | 4  | -1.00   | 0.3739     |
|     | Satterthwaite| Unequal   | 2  | -1.00   | 0.4226     |
| Q40 | Pooled       | Equal     | 4  | -5.00   | 0.0750     |
|     | Satterthwaite| Unequal   | 2  | -5.00   | 0.0577     |
| A40 | Pooled       | Equal     | 4  | -1.75   | 0.1550     |
|     | Satterthwaite| Unequal   | 2  | -1.75   | 0.2222     |

## VII. Statistical treatment of data presented in Table 3

Table SM12 - Nonlinear OLS Summary of Residual Errors to Interfacial Tension Model adjusted to experimental data

| | DF Model | DF Error | MSE | Adjusted R-Square |
|---|---|---|---|---|
| 0.000 g·100 mL$^{-1}$ chitosan + 0.0 mL·(100 mL)$^{-1}$ Tween 20 in lactic acid solution (pH 3.0) | - | - | - | - |
| 0.000 g·100 mL$^{-1}$ chitosan + 0.1 mL·(100 mL)$^{-1}$ Tween 20 in lactic acid solution (pH 3.0) | 3 | 27 | 0.2945 | 0.9916 |
| 0.025 g·100 mL$^{-1}$ chitosan + 0.1 mL·(100 mL)$^{-1}$ Tween 20 in lactic acid solution (pH 3.0) | 3 | 27 | 0.8619 | 0.9615 |
| 0.050 g·100 mL$^{-1}$ chitosan + 0.1 mL·(100 mL)$^{-1}$ Tween 20 in lactic acid solution (pH 3.0) | 3 | 27 | 0.1854 | 0.9983 |
| 0.100 g·100 mL$^{-1}$ chitosan+ 0.1 mL·(100 mL)$^{-1}$ Tween 20 in lactic acid solution (pH 3.0) | 3 | 27 | 0.1603 | 0.9937 |
| 0.000 g·100 mL$^{-1}$ chitosan + 0.1 mL·(100 mL)$^{-1}$ Tween 20 in water | 3 | 27 | 0.2393 | 0.9779 |

Table SM13 - Nonlinear OLS Parameter Estimates to Interfacial Tension Model adjusted to experimental data

| | Parameter | Approx. Std Err | t Values | Approx. Pr > \|t\| |
|---|---|---|---|---|
| 0.000 g·100 mL$^{-1}$ chitosan + 0.0 mL·(100 mL)$^{-1}$ Tween 20 in lactic acid solution (pH 3.0) | $\sigma_{eq}$ | - | - | - |
| | A | - | - | - |
| | b | - | - | - |
| 0.000 g·100 mL$^{-1}$ chitosan + 0.1 mL·(100 mL)$^{-1}$ Tween 20 in lactic acid solution (pH 3.0) | $\sigma_{eq}$ | 0.4752 | 115.33 | < 0.0001 |
| | A | 0.4034 | 42.32 | < 0.0001 |
| | b | 0.00377 | 15.47 | < 0.0001 |
| 0.025 g·100 mL$^{-1}$ chitosan + 0.1 mL·(100 mL)$^{-1}$ Tween 20 in lactic acid solution (pH 3.0) | $\sigma_{eq}$ | 0.7087 | 81.64 | < 0.0001 |
| | A | 0.8225 | 26.44 | < 0.0001 |
| | b | 0.00843 | 9.94 | < 0.0001 |
| 0.050 g·100 mL$^{-1}$ chitosan + 0.1 mL·(100 mL)$^{-1}$ Tween 20 in lactic acid solution (pH 3.0) | $\sigma_{eq}$ | 0.2767 | 203.79 | < 0.0001 |
| | A | 0.2380 | 98.62 | < 0.0001 |
| | b | 0.00171 | 35.48 | < 0.0001 |
| 0.100 g·100 mL$^{-1}$ chitosan+ 0.1 mL·(100 mL)$^{-1}$ Tween 20 in lactic acid solution (pH 3.0) | $\sigma_{eq}$ | 0.3290 | 175.53 | < 0.0001 |
| | A | 0.3821 | 66.41 | < 0.0001 |
| | b | 0.00336 | 24.96 | < 0.0001 |
| 0.000 g·100 mL$^{-1}$ chitosan + 0.1 mL·(100 mL)$^{-1}$ Tween 20 in water | $\sigma_{eq}$ | 1.5040 | 32.41 | < 0.0001 |
| | A | 1.3334 | 17.30 | < 0.0001 |
| | b | 0.00328 | 9.29 | < 0.0001 |

Table SM14 - ANOVA used to evaluated $\sigma_{eq}$, $A$ and $b$ parameters from Interfacial Tension Model

| Parameter | DF | Mean Square | F Value | Pr > F |
|---|---|---|---|---|
| $\sigma_{eq}$ | 4 | 58.2931731 | 3.70 | 0.04225 |
| $A$ | 4 | 12.38810667 | 3.35 | 0.0551 |
| $B$ | 4 | 0.005760000 | 3.04 | 0.0699 |